\documentstyle[epsf,multicol,aps,prl,tighten]{revtex}
\renewcommand{\narrowtext}{\begin{multicols}{2} 
\global\columnwidth20.5pc} 
\renewcommand{\widetext}{\end{multicols} 
\global\columnwidth42.5pc}   
\def\top#1{\vskip #1\begin{picture}(290,80)(80,500)\thinlines \put(    
65,500){\line( 1, 0){255}}\put(320,500){\line( 0, 1){   
5}}\end{picture}} 
\def\bottom#1{\vskip #1\begin{picture}(290,80)(80,500)\thinlines \put( 
330,500){\line( 1, 0){255}}\put(330,500){\line( 0, -1){ 
5}}\end{picture}} 

\newcommand{\eq}{\begin{equation}}
\newcommand{\ee}{\end{equation}}
\newcommand{\eqa}{\begin{eqnarray}}
\newcommand{\eea}{\end{eqnarray}}

\newcommand{\tg}{{\widetilde\Gamma}}
\newcommand{\vp}{{\vec{p}}}
\newcommand{\vk}{{\vec{k}}}
\newcommand{\vq}{{\vec{q}}}
\newcommand{\bb}{{\tilde\beta}}
\begin{document}
\draft
\title{Superconductivity near Itinerant Ferromagnetic Quantum Criticality
}
\author{Ziqiang Wang$^a$, Wenjin Mao$^{a,b}$, and Kevin Bedell$^a$}
\address{$^a$Department of Physics, Boston College, Chestnut Hill, MA 02467}
\address{$^b$Department of Physics, Rutgers University, 
Piscataway, NJ 08854}
\date{\today}
\maketitle
\begin{abstract}
Superconductivity mediated by spin fluctuations in weak 
and nearly ferromagnetic metals is studied close to the zero-temperature
magnetic transition. 
We solve analytically the Eliashberg equations for $p$-wave
pairing and obtain the normal state quasiparticle self-energy 
and the superconducting transition temperature $T_c$
as a function of the distance to the quantum critical point.
We show that the reduction of quasiparticle coherence and life-time due
to scattering by quasistatic spin fluctuations
is the dominant pair-breaking process, which leads to a rapid
suppression of $T_c$ to a nonzero value near the quantum critical point. 
We point out the differences and the similarities of the problem to that of
the theory of superconductivity in the presence of
paramagnetic impurities.
\end{abstract}
\narrowtext
Superconducting ground states have recently been discovered in materials
sitting close to the zero-temperature phase boundary of both 
antiferromagnetic and ferromagnetic transitions 
\cite{exptaf,exptfm}. It is quite natural to suspect that paring in these 
materials is mediated by spin-fluctuations which are enhanced near 
the magnetic quantum critical point (QCP) 
\cite{fisk,coleman,singh,mont,chub,roussev,kb}.
However, it is known that quasiparticles scattered by critical
spin-fluctuations are subject to severe non-Fermi liquid 
self-energy corrections that are in general pair-breaking.
The interplay between these two competing effects, generic to
superconductivity near quantum phase transitions, has not been
fully understood and is the subject of this work.

We focus on the case of p-wave superconductivity near
the ferromagnetic QCP where the Curie temperature
is driven to zero by, {\it e.g.} pressure or doping \cite{exptfm,MnSi}.
In the vicinity of the transition, the metals can
be described by weak ferromagnetic (FM) and nearly FM Fermi liquids
at low temperatures, respectively \cite{dk,doniach}. 
The important low energy excitations are the quasiparticles and
the {\it long wavelength} spin fluctuations.
Since the quasiparticles couple strongly to spin fluctuations near the QCP,
it is necessary to adapt the framework of the 
strong coupling Eliashberg equations.
The long wavelength nature of the
critical spin fluctuations near the ferromagnetic QCP \cite{millis} makes the
solution of these equations theoretically tractable, enabling us to
obtain analytical results for the self-energy and the
transition temperature $T_c$. We find the onset of p-wave
superconductivity in both the Fermi liquid $T_c<T^*$
and the quantum critical regime $T_c> T^*$, where $T^*$ is
the characteristic frequency for spin fluctuations.
We show that $T_c$ is rapidly reduced on 
approaching quantum criticality with $T^*\to0$ due to
the rapid reduction of quasiparticle coherence and life-time
caused by quasistatic scattering of spin fluctuations
analogous to the suppression of $T_c$ by paramagnetic impurities \cite{ag}.
Interestingly, $T_c$ remains finite at the FM QCP.
We shall not discuss the problem of coexistence of
ferromagnetism and superconductivity here, because the latter requires
a careful treatment of the coupling between the superconducting order 
parameter and the electromagnetic fields associated with the
ferromagnetic fluctuations \cite{varma} that is beyond the
scope of this work.

In a FM Fermi liquid with spin polarization along the $z$-axis,
the single-particle Green's function has the form,
$
G_{\sigma}(\epsilon,{{\vp}})=a_\sigma/[\epsilon-v_\sigma
(\vp-p_\sigma)+i\eta {\rm sgn} (\epsilon)],
$
where $\eta \rightarrow 0^+$, $p_\sigma$ ($\sigma=\uparrow,\downarrow$)
are the Fermi momenta of the spin up and down electrons,
and $a_\sigma$ are the wave-function renormalizations.
For a weak FM metal, such as the one close to a continuous
FM transition, the difference in the Fermi momenta is small,  
$\delta\equiv\vert p_\uparrow-p_\downarrow\vert\ll p_{\uparrow,\downarrow}$,
and it is possible to set $p_\uparrow=p_\downarrow=p_F$ and 
$a_\uparrow=a_\downarrow$ \cite{dk}.
The spin fluctuations are described by the propagators of the
electron spin-density,
$D_{ij}({\vec x},\tau)=-i\langle(S_i({\vec x},\tau)-\langle S_i\rangle)
(S_j(0,0)-\langle S_j\rangle)\rangle$, ${i,j}=x,y,z$.
The long wavelength and low energy spin-fluctuations in the transverse 
and longitudinal channels are given by \cite{dk},
\begin{eqnarray}
{D}_{\parallel} (\omega,{{\vq}})&=&
-{N_F\over2}
{1\over \alpha+({q\over 2p_F})^2-i{\pi\omega\over4\Lambda}{2p_F\over q}},
\label{dpara} \\
D_\perp (\omega,\vq)&=& \left\{ \begin{array}{ll}
-\displaystyle{{N_F\over2}
{1\over ({q\over2p_F})^2- i{\pi\omega\over4\Lambda}{2p_F\over q}}}, 
& q \gg |\delta|,
\\
-\displaystyle{S}{\omega_s ({{q}})\over
\omega^2-\omega_s ^2 ({{q}})}, & q \ll |\delta|.
\end{array} \right.
\label{dperp}
\end{eqnarray}
Here $\alpha\approx (\delta/p_F)^2\ll1$ measures the distance to the
critical point, $S=v_F \delta N_F/2$ is the averaged uniform spin density,
$\Lambda=2 v_F p_F$ is an energy scale of the
order of the Fermi energy, and $N_F\simeq p_F^2/(2\pi^2 v_F)$
is the density of states per spin at the Fermi level.
Notice that the transverse spin-wave only
emerges for $q\ll\delta$ with a dispersion
$\omega_s(q)\simeq v_F\delta(q/2p_F)^2$.

On the paramagnetic (PM) side, the spin rotation symmetry is restored. 
The spin fluctuations become isotropic, $D_\perp =D_\parallel\equiv D$.
All three modes take on the paramagnon form in Eqs.~(\ref{dpara}) with
$\alpha$ determined by the spin correlation length $\xi_s$,
$\alpha\sim\xi_s^{-2}$. 

FM spin fluctuations are known to be pair-breaking
in the s-wave channel and
therefore suppress the conventional phonon-mediated
superconductivity in PM metals. This problem
was studied by Berk and Schrieffer \cite{berk}
using the strong coupling Eliashberg theory
and is believed to be the reason
for why transition metals close to the FM instability have, if any,
a low $T_c$. In contrast, the pairing interaction
is attractive in the $l={\rm odd}$ angular momentum channel, raising 
the interesting possibility of FM spin fluctuation mediated 
spin-triplet superconductivity \cite{fay}.
In general, the presence of a 
spontaneous magnetization in the FM phase leads to a set
of four Eliashberg equations. However, for weak ferromagnets with small
moments, the spin dependence of the self energy and the gap function
can be ignored.
For notational convenience, we shall limit the presentation
to the PM phase, the modification on the
FM side is straightforward and the differences will be noted 
explicitly. 

Expressing the self energy in the Nambu formalism,
$\Sigma_p(\omega)\equiv [1-Z_p(\omega)]
\omega\tau_0+Z_p(\omega)\Delta_p(\omega)\tau_1$, where $\Delta_p(\omega)$
is the gap function, the linearized Eliashberg equations 
for the self-energy and the gap function are given by \cite{doug},
\begin{eqnarray}
i\omega_n[1-Z_p(i\omega_n)]&=& -g_0^2 s_0T\sum_{i\epsilon_n}\int
{d^3k\over(2\pi)^3}
{1\over i\epsilon_n Z_k(i\epsilon_n)-\xi_k}
\nonumber \\
&&
\quad\times D(\vp-\vk,i\omega_n-i\epsilon_n),
\label{selfenergy} \\
Z_p(i\omega_n)\Delta_p(i\omega_n)&=&g_0^2 s_lT\sum_{i\epsilon_n}\int
{d^3k\over(2\pi)^3}
{Z_k(i\epsilon_n)\Delta_k(i\epsilon_n)\over
\epsilon_n^2Z_k^2(i\epsilon_n)+\xi_k^2}
\nonumber \\
&& \quad
\times D(\vp-\vk,i\omega_n-i\epsilon_n).
\label{pairing}
\end{eqnarray}
Here, $g_0$ is the coupling constant of the quasiparticles to
spin fluctuations and $\xi_k=k^2/2m-\mu$.
Note that the presence of Heisenberg symmetry in the PM phase
guarantees three identical soft modes contributing to
the self-energy but one longitudinal mode to the gap equation 
for triplet (p-wave) pairing \cite{mont}, i.e. $s_0=3$ and $s_1=1$.
To study the influence of the departure from this symmetry on $T_c$,
such as in the case of Ising spins with $s_0=s_1=1$,
we keep $s_0$ and $s_l$ as general parameters \cite{roussev}.
On the FM side, spin rotation symmetry is broken.
While the gap equation (\ref{pairing}) only contains the longitudinal 
mode and stays invariant,
the self-energy equation (\ref{selfenergy}) needs to be modified in a 
straightforward manner to
reflect the different contributions from the transverse modes.

It is customary to proceed by
separating the $k$-integral into angular and amplitude parts,
$
\int d^3k\to (4p_F^2/v_F)\int x dx
\int d\phi\int d\xi_k,
$
where $x=\sin(\theta/2)$, $\theta$ is the angle between ${\vk}$
and 
${\vq}={\vp}-{\vk}$, and $\xi_k$ satisfies the kinematic constraint 
$q^2\simeq4p_F^2x^2+\xi_k^2/v_F^2$.
As it will turn out later, close to the QCP and
compared to their frequency dependence, the
self energies have a negligibly weak momentum dependence. We thus drop
the $k$-dependence in $Z_k(i\omega)\to Z(i\omega)$ and project the gap 
function into the $l$-th angular momentum channel, 
$\Delta_k(i\omega)\to\Delta_l(i\omega)$.
With this approximation, it is possible to carry out the integral over
$\xi_k$.
In order to make analytical progress, we analytically continue
to real frequencies and obtain,
\widetext
\top{-2.8cm}
\begin{eqnarray} 
[1-Z(\omega)]\omega&=&-s_0g^2\int_{-\infty}^{\infty}d\epsilon
\int_0^1 xdx\biggl[ 
\bigl({1\over U(\omega-\epsilon)[U(\omega-\epsilon)-i\epsilon 
Z(\epsilon)/\Lambda]}
+(iZ\to-iZ^*)\bigr) 
\tanh({\epsilon\over2T}) 
\nonumber \\ 
&& \qquad\qquad 
+\bigl({1\over U(\epsilon)[U(\epsilon)-i(\omega-\epsilon) Z(\omega-\epsilon)/
\Lambda]}
-(U\to U^*)\bigr) 
\coth({\epsilon\over2T})\biggr], 
\label{realenergy} \\ 
Z(\omega)\Delta_l(\omega) &=& s_lg^2\int_{-\infty}^{\infty}d\epsilon 
\int_0^1 xdx \biggl[ 
\bigl({\Delta_l(\epsilon)\over\epsilon}{1\over U(\omega-\epsilon)
[U(\omega-\epsilon) -i\epsilon Z(\epsilon)/\Lambda]}+ 
(iZ,\Delta_l\to-iZ^*,\Delta_l^*)\bigr)\tanh({\epsilon\over2T})  
\nonumber \\ 
&&\qquad\qquad+\bigl({\Delta_l(\omega-\epsilon)\over\omega-\epsilon}{1\over  
U(\epsilon)[U(\epsilon)-i(\omega-\epsilon) Z(\omega-\epsilon)/\Lambda]} 
-(U\to U^*)\bigr)\coth({\epsilon\over2T})\biggr]P_l(1-2x^2), 
\label{realgap} 
\end{eqnarray} 
\bottom{-2.7cm} 
\narrowtext 
\noindent where $g^2=g_0^2N_F^2/2$, $P_l(x)$ is the Legendre polynomial,
and $U^2(\epsilon)=\alpha+x^2-i\pi\epsilon/4x\Lambda$.

We first solve Eq.~(\ref{realenergy}) to derive the quasiparticle 
self energy in the normal state.
The self consistency in the this equation is crucial near the QCP, since as
$\alpha\to0$, the important low frequency cutoff in the denominators
is the self energy itself. 
Carrying out the integrals, it is straightforward to show that
the dominant contributions to the self energy come from
the scattering by the spin fluctuations with momentum transfer $q\gg \delta$
and having the same form in both the PM and the FM phase close to the QCP.
We find that the characteristic energy scale for spin fluctuation,
$
T^*\sim \alpha^{3/2}\Lambda,
$
enters as an important crossover temperature scale.
For $y={\rm max}(T,\epsilon) < T^*$, 
the self energy behaves as in a Fermi liquid,
\eq
\Sigma(\epsilon,T)\approx
-c^\prime\epsilon\ln(\Lambda/T^*)-ic^{\prime\prime} y^2/T^*, \quad y< T^*,
\label{sigfermi}
\ee
where $c^\prime,c^{\prime\prime}\sim s_0g^2$.
However, for $y> T^*$, the scattering 
by spin fluctuations is enhanced and
the self energy becomes non Fermi liquid like with
the real part,
\eq
\Sigma^\prime(\epsilon,T)\approx -c^\prime\epsilon\ln (\Lambda/y), 
\quad y>T^*.
\label{sigre}
\ee
This leads to a quasiparticle residue ${\cal Z}$,
$
{\cal Z}^{-1} = 1-[\partial \Sigma^\prime/\partial 
\epsilon]\vert_{\epsilon=0}=1+c^\prime\ln[\Lambda/{\rm Max}(T^*,T)],
$
that vanishes logarithmically on approaching the QCP
as in a marginal Fermi liquid \cite{marginal}.
For $\epsilon> T\gg T^*$, the imaginary part of $\Sigma$ follows,
\eq
\Sigma^{\prime\prime}(\epsilon)\approx -c^{\prime\prime}{\pi\epsilon/2},\quad
\epsilon> T\gg T^*.
\label{sigim1}
\ee
In the quasistatic regime,
$T >\epsilon\gg T^*$, another energy scale arises by comparing
the coherence length $\xi_{\rm coh}\sim v_F/T$ to the spin
correlation length $\xi_s\sim 1/\sqrt{\alpha}$.
We find for $\xi_{\rm coh} > \xi_s$,
\eq
\Sigma^{\prime\prime}(T)\approx -c^{\prime\prime} T\ln{T\over T^*},\quad
{T\over\Lambda}\ln{\Lambda\over T}\ll ({T^*\over \Lambda})^{1/3}.
\label{sigim2}
\ee
For even smaller $T^*$, such that $\xi_{\rm coh} < \xi_s$, the quasiparticles
scatter off essentially uncorrelated spins. In this case, the unphysical
singularity in Eq.~(\ref{sigim2}) as $T^*\to0$ must be removed by 
the self-consistency of Eq.~(\ref{realenergy})
where $\Sigma^{\prime\prime}$ itself becomes the cutoff. We find
\eq
\Sigma^{\prime\prime}\approx -c^{\prime\prime}T\ln(\Lambda^2T/\vert
\Sigma^{\prime\prime}\vert^3)+3c^{\prime\prime}\sqrt{\alpha}T\Lambda/2
\vert\Sigma^{\prime\prime}\vert,
\label{sigim3}
\ee
which has the following self-consistent solution,
\eq
\Sigma^{\prime\prime}\approx -c^{\prime\prime}T\ln{\Lambda\over T}
+{3\over2}c^{\prime\prime}\sqrt{\alpha}\Lambda,\quad
{T\over\Lambda}\ln{\Lambda\over T}\gg ({T^*\over \Lambda})^{1/3}.
\label{sigim4}
\ee
The same behavior of $\Sigma$ holds on the FM side since the dominant
scattering process has $q\gg\delta$.
In this range of $q$, the breaking of spin rotation invariance
is insignificant as seen in Eqs.~(\ref{dpara}) and (\ref{dperp}).
The spin wave contribution is
not important since the difference between the
Fermi momentum ($\delta$) is much larger than 
most of the momenta carried by the spin waves.
Notice that the inelastic scattering rate in quasistatic limit 
increases with a square-root singularity as 
$\alpha\to0$, leading eminently to a rapid suppression of the superconducting 
transition temperature $T_c$ on approaching the QCP.

We next determine $T_c$ by solving Eqs.~(\ref{realgap}) 
for the simplest $l=1$, p-wave case \cite{notepwave}
To treat the effects of both mass renormalization
and scattering lifetime, we write $Z=Z'+iZ^{\prime\prime}$ and the
complex gap function as $\Delta=\Delta^\prime
+i\Delta^{\prime\prime}$.
Taking the imaginary part of the gap equation (\ref{realgap}), we obtain
to leading order in $T_c/T_0$, $T_0\sim\Lambda$ being the cutoff
frequency for spin fluctuations - a magnetic analog of Debye frequency,
\eq
Z^\prime(\omega)\Delta^{\prime\prime}(\omega)+\beta
Z^{\prime\prime}(\omega)\Delta^\prime(\omega)s_l/s_0\simeq 0.
\label{deltare}
\ee
Here $\beta=s_0/s_l-1$ reflects the spin symmetry.
Writing for small $\omega$,
$
\omega Z(\omega)=[1+\lambda(T)]\omega-i \Gamma(\omega),
$
where $\lambda=-\Sigma^\prime/\omega$
is the effective coupling and $\Gamma=-\Sigma^{\prime\prime}$ 
is half the inverse life-time of the quasiparticles, 
Eq.~(\ref{deltare}) becomes
$
\beta\Gamma(\omega)\Delta^\prime(\omega)=
(1+\lambda)\omega\Delta^{\prime\prime}s_0/s_l,
$
which allows us to account for the damping
of the order parameter in terms of a real effective gap function,
\eq
\Delta_{\rm eff}(\omega)=\Delta^\prime(\omega)[1+(\tg/\omega)^2],
\label{egap}
\ee
where $\tg=\beta s_l\Gamma/s_0(1+\lambda)$. In contrast to
$\Delta^\prime$, $\Delta_{\rm eff}$ has a weaker 
$\omega$-dependence and remains finite in the small $\omega$ limit.
Now we can rewrite the real part of the gap equation (\ref{realgap}) as
an integral equation for $\Delta_{\rm eff}$,
\begin{eqnarray}
(1+\lambda)&&\Delta_{\rm eff}(\omega)
=2s_l g^2 \int_{-\infty}^{\infty} d \epsilon\int_0^1 xdx P_l(1-2x^2)
\nonumber \\
&&\tanh({\epsilon\over2T_c}){\epsilon\Delta_{\rm eff}(\epsilon)\over
\epsilon^2+\tg^2(\epsilon,T_c)}{\rm Re}{1\over U^2(\omega-\epsilon,x)}.
\label{effgap}
\end{eqnarray}
The scattering rate enters as a low-energy cutoff of the
logarithmic singularity responsible for the superconducting
instability.
Next we attempt an approximate analytical solution of $T_c$
from Eq.~(\ref{effgap}) where the
weak frequency dependence of $\Delta_{\rm eff}$ can be neglected.

{\sl Fermi liquid regime}. Consider first the case 
$T_c \ll T^*\ll T_0$, i.e. the onset of superconductivity in the 
Fermi liquid regime away from the QCP. Since the temperature is much
lower than the characteristic spin fluctuation frequency, inelastic
scattering dominates, but with the ordinary
Fermi liquid scattering rate (see Eq.~\ref{sigfermi}) that is much 
smaller than max$(k_B T,\epsilon)$. The effect of a
nonzero $\tg$ on $T_c$ is thus small and negligible.
Solving Eq.~(\ref{effgap}) to next to leading order in $T^*/T_0$,
we obtain
\eq
T_c\simeq T_0 e^{-[\eta^\prime+\beta(2A+3)+A^2]/(2A-3)},
\label{tcfermiliquid}
\ee
where $A=\ln T_0/T^*$, $\eta^\prime=(3/2s_lg^2)+(\pi^2/24)
+6\ln(\sqrt{2}\gamma/\pi)-3$, and 
$\ln\gamma\simeq0.577$ is Euler's constant.
As $T^*$ is reduced towards the QCP,
spin fluctuations increase and pairing is enhanced. 
This causes $T_c$ to rise initially.
Reducing $T^*$ further eventually 
causes the system near $T_c$ to lose
sensitivity to the finite correlation length, 
leading to new physics associated with
superconductivity near quantum criticality.

{\sl Quantum critical region}. Here, $ T_0 \gg T_c \gg T^*$,  the
superconducting transition occurs inside the quantum critical
regime.
Since the temperature is much higher than the characteristic
quantum spin fluctuation energy, inelastic
scattering is negligible and the dominant 
pair-breaking effect comes from quasistatic 
($\omega < T_c$) spin fluctuations with a 
scattering rate
$
\tg(T)=-\beta s_l\Sigma^{\prime\prime}(T)/s_0[1+\lambda(T)].
$
The suppression of $T_c$ due to $\tg$ is thus reminiscent of 
Abrikosov and Gor'kov's theory of
superconducting alloys with paramagnetic impurities \cite{ag}.
Accordingly, 
Eq.~(\ref{effgap}) has the solution,
\eq
\ln{T_c\over T_{c0}}=\psi\left({1\over2}\right)-\psi\left[{1\over2}
+{\tg(T_c)\over 2\pi T_c}\right],
\label{tc}
\ee
where $\psi$ is the digamma function and $T_{c0}$ is the transition
temperature in the absence of $\tg$. To leading order in $Tc/T_0$,
\eq
{T_{c0}\over T_0}=e^{-[\bb+\sqrt{\bb^2+2\beta[1-{\pi\over\sqrt{3}}
({T^*\over T_{c0}})^{2/3}]+\eta+{\pi\over\sqrt{3}}({T^*\over T_{c0}})^{2/3}}]},
\label{tc0}
\ee
with $\bb=3+\beta$ and $\eta=(3/2s_lg^2)+6\ln(2\gamma/\pi)-\pi^2/24$.
At the QCP, we find $T_{c0}/T_0 \sim 10^{-5}$ for Heisenberg symmetry 
and $\sim 10^{-3}$ for Ising symmetry.

A few remarks are in order for $T_c$ in the quantum
critical regime.
(i) For Ising spins, $\beta=0$. Eq.~(\ref{tc}) shows that 
$T_c$ is not affected by a finite quasiparticle life-time,
i.e. $T_c\simeq T_{c0}$. Furthermore, to leading order, $T_{c0}$ is not reduced
by the real part of the self energy.
This is in fact a manifestation of Anderson's theorem \cite{anderson}
for nonmagnetic impurities. It arises in our case from the cancellation of
the self-energy effects in the gap equation (\ref{effgap})
in the quasistatic limit when $s_0=s_l$.
From Eq.~(\ref{tc0}), it follows that $T_c$ decreases linearly
with $\alpha$ close to the QCP with a slope $dT_c/d\alpha
\sim -T_c(T_0/T_c)^{2/3}$ for Ising spins.

(ii) For Heisenberg spins,  $\beta=2$. $T_{c0}$ decreases with the
reduction of quasiparticle coherence on approaching the QCP.
$T_c$ is further reduced from $T_{c0}$ due to the increasing scattering rate.
However, in contrast to the case of magnetic impurities where $T_c$ can be
suppressed to zero at a finite concentration, we find that
{\it $T_c$  remains finite at the QCP} as a consequence of
the $T$-dependent scattering rate. From
Eqs.~(\ref{sigre},\ref{sigim4}) at $\alpha=0$, it follows that
$
{\tg(T_c)/ T_c}
=\beta s_l c^{\prime\prime}\ln (\Lambda/T_c)/
s_0[1+c^{\prime} \ln (\Lambda/T_c)]
$
tends to a constant $\rho=c^{\prime\prime}\beta s_l/2s_0\pi c^{\prime}$
of order unity for $T_c\ll \Lambda$, leading to
$T_c\simeq T_{c0}\exp-[\psi(1/2+\rho)-\psi(1/2)]$.
\begin{figure}      
\vspace{-0.5truecm}  
\center      
\centerline{\epsfysize=2.8in      
\epsfbox{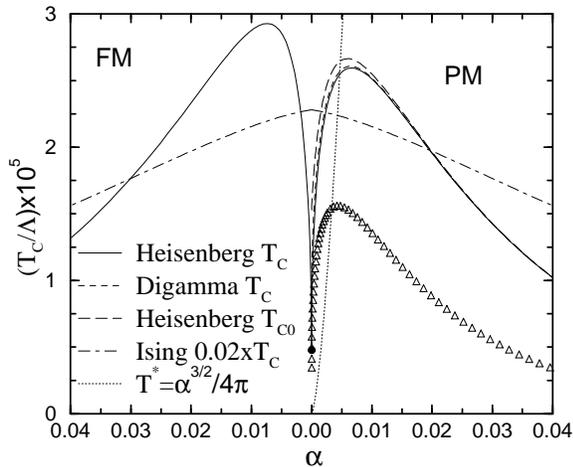}}  
\vspace{-0.5truecm}      
\begin{minipage}[t]{8.1cm}      
\caption{ $T_c$ versus $\alpha$ for $g^2=0.3$ (lines) and $0.15$ (triangles).
$T_{c0}$ and $T_c$ obtained from 
Eq.~(\ref{tc}) are shown for comparison. The peak in $T_c$ versus $\alpha$
scales with $T^*$.
}    
\label{fig1}     
\end{minipage}      
\end{figure}      
The results obtained from numerical solution
of Eq.~(\ref{effgap}) are shown in Fig.~1. On the PM side,
they are in agreement with those of Roussev and Millis \cite{roussev}.
We next analyze how $T_c$ varies with $\alpha$ close to the QCP in the
Heisenberg case. Eq.~(\ref{tc0}) shows that $T_{c0}$ increases linearly
with $\alpha$, as the overall sign of the terms proportional to 
$(T^*/T_c)^{2/3}$ has changed from the Ising case due to the real part of
the self energy when $\beta=2$. 
However, we find that this effect
is subleading, and the dominant $\alpha$-dependence of $T_c$ 
comes from the scattering rate in
Eq.~(\ref{tc}) through the strong $\alpha$-dependence
of $\Sigma^{\prime\prime}$ near the QCP, 
i.e. $d T_c/d\alpha\propto -d\Sigma^{\prime\prime}/d\alpha$.
From Eqs.~(\ref{sigim2},\ref{sigim4}), we obtain,
$dT_c/d\alpha\sim 1/\sqrt{\alpha}$, 
for $\sqrt{\alpha}\ll (T_c/\Lambda)\ln(\Lambda/T_c)$;
and 
$
dT_c/d\alpha\sim 1 /\alpha
$, 
for $(T_c/\Lambda)\ln(\Lambda/T_c)<\sqrt{\alpha} < (T_c/\Lambda)^{1/3}$.

In the FM phase, the spin rotation symmetry is broken.  However, 
close to the transition, approximate Heisenberg symmetry is restored due 
to the small difference in the Fermi momenta which in turn, as discussed above,
suppresses the contributions of the long wavelength 
Goldstone mode to the electron-spin fluctuation kernel. 
As a result, the superconducting phase boundary
is approximately symmetric near the magnetic QCP. Away from
the QCP, the deviation from the Heisenberg symmetry leads effectively to a
$\beta$ value that is shifted downward from the Heisenberg value and
a somewhat higher $T_c$ (see Fig.~1) on the FM side. 
Well inside the FM phase, the relative suppression of the 
fluctuation in the transverse channel
makes the situation closer to the Ising case 
studied above, resulting in a much higher $T_c$ \cite{exptfm}.

To conclude, our results suggest whether a significant suppression
of $T_c$ occurs near the QPC can be used to help identify the spin symmetry of
the superconducting order parameter. We have shown that such a reduction
occurs in the triplet case but is absent for singlet pairing, e.g. the
s-wave pairing proposed in the weak ferromagnetic {\it local} Fermi liquid
theory \cite{kb}. In the singlet case, 
the spin fluctuations contributing to 
the self energy and the gap equation are identical and the dominant quasistatic
pair-breaking effects in the quantum critical regime
cancel out as in the Ising case discussed above. While more experiments are
clearly needed to determine the phase diagram near the FM QCP \cite{exptfm},
existing data in the AF case \cite{exptaf} show that $T_c$ 
indeed peaks near the magnetic QCP where pairing due to AF spin 
fluctuations is expected to be of spin-singlet in nature.

This work is supported in part by
DOE Grant Nos. DE-FG02-99ER45747, DE-FG02-97Er45636,
and by an award from Research Corporation. 
 
\end{multicols} 

\begin{references}
\bibitem{notepwave} Since the pairing interaction near the QCP
is isotropic to leading order in the magnetization, the three
components of the p-wave order parameter are degenerate.
\end{references}
\end{document}